# FlexKeys: Rapidly Customizable 3D Printed Tactile Input Devices with No Assembly Required

*Ben Greenspan[1], Eric M Gallo[1], Andreea Danielescu[1]*

[1] Accenture Labs
San Francisco, CA 94105, USA

## ABSTRACT

Physical input devices serve as a tactile interface between users and computing systems. These devices are often complex assemblies that consist of both electrical and mechanical components making customization difficult and out of reach for non-engineers. While these components can now be 3D printed on demand, they must still be independently designed and assembled. We present FlexKeys, an approach in which devices that include both electrical and deformable components can be created in a single print on a multi-material 3D printer, requiring no assembly. Designers can customize devices including the input type, travel distance and layout of keys, textures of surfaces, and route all electrical signals directly to a microcontroller socket. In many instances, these devices require no support material, producing a functional device the moment a print finishes. We demonstrate this approach by creating a customized keyboard and report on validation measurements of individual input keys as well as highlighting additional designs. This work provides the first step towards lowering the barrier to entry for non-engineers to design custom tactile inputs, enabling occupational and physical therapists, clinicians, and educators to design and create devices directly based on their assessments of individual user needs.

# INTRODUCTION

The design and manufacturing of physical input devices, such as keyboards, is a complex process requiring the design, manufacturing, electrical wiring, and assembly of several components such as buttons, keys, switches, and triggers. Even a single component, like a key, is complex, composed of a deformable spring, electronic traces and a rigid key cap or housing at minimum. Each component is often designed in separate computer aided design (CAD) tools, requires its own manufacturing process, and then must be assembled into a functional device (Customs, 2019). The number of steps, tools, and processes necessary to create a single device adds to the overall complexity, requiring multiple designers and engineers. Because of this, making small changes to a design, like customizing the layout of a device to match individual user needs, is challenging, time intensive, and often not cost effective.

3D printing provides an ideal alternative to traditional manufacturing for production of low volume, complex structures, but to date, has primarily been used for prototyping static parts (Fuge *et al.*, 2015; Sosin, 2019; Printing, 2021). New techniques now enable motion or deformation within printed parts without the need for assembly by modifying the thickness of specific areas and tuning printing properties such as the number of shells, infill percentage and pattern. By exploiting these properties, we can create 3D printed springs (He *et al.*, 2019), flexible mechanisms that achieve motion through elastic deformation (Howell, 2013; Megaro *et al.*, 2017) and materials that derive their properties from their structure rather than the material they are made from (Schumacher *et al.*, 2015; Ion *et al.*, 2016, 2017; Gong *et al.*, 2021). All of these have potential applications for physical input devices. Additionally, conductive composite filaments can be created by mixing conductive material with traditional printable materials, allowing for traces, resistors, inductors, capacitors and filters to be 3D printed (Flowers *et al.*, 2017). Structural electronics can be created by printing multiple materials at once (Lazarus and Tsang, 2020) to produce components such as antennas, buttons, knobs, and sliders (Burstyn *et al.*, 2015; Iyer, Chan and Gollakota, 2017; Gong *et al.*, 2021), coil spring strain sensors (Greenspan and Danielescu, 2020), and capacitive touch interfaces (Burstyn *et al.*, 2015; Schmitz *et al.*, 2015; Götzelmann and Schneider, 2016; Takada, Shizuki and Tanaka, 2016; Davis *et al.*, 2020). To use these new materials and capabilities, plugins have been created within non-parametric CAD programs, such as Grasshopper within Rhinoceros. While these plugins provide custom visual interfaces that help experts leverage new materials and capabilities, applying multiple materials is still challenging and these tools are not easily accessible to non-experts.

We present FlexKeys, an approach to design and create tactile input devices that leverages existing design software and accessible 3D printing technology to reduce the number manufacturing steps and eliminate the need for assembly. The process enables a designer to create a completely customizable physical input device using a library that includes keys with multiple input types, travel distances, activation forces and keycap textures. The result is a fully connected tactile input device created in a single print, requiring only the insertion of a microcontroller board to complete. We present the design and evaluation of the mechanical performance of a custom keyboard as an example input device and highlight additional applications. We conclude with a discussion of how this design approach provides the first step towards enabling individuals with no engineering background to design and create custom physical input devices.

## CUSTOMIZABLE DESIGN APPROACH

All devices are designed in a single parametric CAD program to allow customization of each key. We chose Autodesk Fusion 360 because it's the least expensive and most accessible among professional CAD software suites. Devices were sliced in a free software, Ultimaker Cura, and printed on a fused deposition modeling (FDM) Ultimaker S5 printer with two print heads. FDM was chosen for its multi-material functionality and because it's the most popular 3D printing method (Printing, 2021). Print head one extruded standard polylactic acid (PLA) filament from Ultimaker and print head two extruded ProtoPasta Conductive Composite PLA (cPLA) filament. This cPLA was selected because it can be printed at a similar temperature to traditional PLA and doesn't require a specialty nozzle capable of higher temperatures or a larger diameter to print. We created a base digital and analog input design that can be produced in a single print and expanded these designs to create a library of parts including keys with multiple travel distances, activation forces, and keycap textures.

### DESIGNING 3D PRINTED INPUT KEYS

We designed two types of inputs based on keys commonly found in input devices: digital (binary) and analog (continuously varying signal). We concentrated on creating deformable, key type inputs, though other input styles, such as sliders, knobs, or dials can also be designed using this same method. Each key, and all their possible configurations, were then saved as independent part files in a master library.

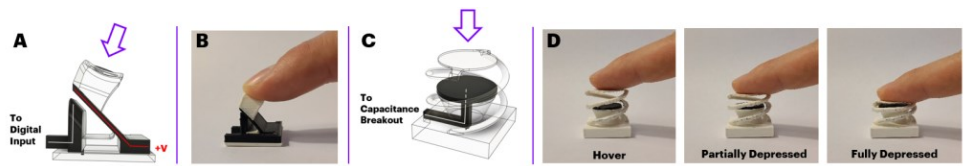

Figure 1. A) Digital key electrical path B) Pressing digital key C) Analog key electrical path D) Analog key hover, partially depressed and fully depressed positions

### Digital Key

The digital key is made from two parts, a conductive cantilever spring that deforms under the application of force, and a non-conductive keycap and base. The conductive part is made from two bodies, one rigid and acts as the return electrode, connected to the digital input of a microcontroller, and the other a cantilever spring which acts as the signal electrode. A constant voltage is sent through the signal electrode such that when no force is applied the circuit is open (Figure 1A). When the key is depressed, the cantilever spring bends to make contact with the return electrode, closing the circuit and registering a keystroke (Figure 1B). Cantilever springs were printed at a 45° angle to eliminate the need for support material. Each conductive part has a complementary non-conductive part made from two bodies, a base and a keycap. The design of the digital key can be further customized for travel distance, activation force and keycap texture.

*Travel Distance:* The digital key we designed supports a wide range of travel distances, accommodating individual user needs and preferences. Travel distance is controlled by modifying the distance between the cantilever spring and the return electrode. We created a set of keys with a short (0.5 mm), medium (1.0 mm), and long (1.5 mm) travel distance based on the Apple MacBook Pro keyboard (travel distance between 0.7 mm to 1.0 mm) (Chokkattu, 2019).

*Activation Force:* Individuals have varying activation force requirements and preferences. The necessary activation force to complete a keystroke can be adjusted by modifying the thickness of the cantilever spring. We created two spring stiffness values utilizing design for manufacturing (DFM) principals where thickness is derived from the width of the actual printed lines. The minimum individual printed line thickness is limited by the 0.4 mm standard diameter of a FDM print head nozzle. We chose a thickness of three printed lines (1.2 mm) for the low stiffness (low activation force), and four lines (1.6 mm) for a high stiffness (high activation force). The three printed lines will withstand the repeated bending cycles the cantilever spring will experience. At the 45° print angle, the resultant thickness of the cantilever spring is 0.85 mm and 1.13 mm respectively.

*Base and Keycap:* The keycap's texture is customizable to create any indicator (letters, numbers, symbols, image, braille) using a single simple extrude. For this example, we designed the keycap width to be 15.5 mm to mimic the average keycap width on a standard keyboard (Cabrera, 2021). The base provides the necessary support for the cantilever spring and default key spacing. Joining the base of one key to an adjacent key creates a spacing of 3.3 mm, also based on the key spacing of a standard keyboard (Cabrera, 2021). The conductive trace that connects to the positive electrode is designed to extend to the left and right ridge of the base, creating a continuous conductive row between keys, simplifying the application of signal voltages.

**Analog Key**

The analog key is also made from two parts, a non-conductive part with a base, a coil spring, a keycap, and a conductive flat circular electrode. The flat circular electrode extends to the bottom of the key and is read by a capacitance breakout board, such as the Adafruit MPR121, and attached to a microcontroller (Figure 1C). Key deformation is detected when a user's finger compresses the spring and moves closer to the electrode beneath it, acting as a parallel capacitor. We validated that the geometry is sensitive enough to detect when a user's finger is hovering on top of the key, partially deformed, and fully deformed (Figure 1D). Our design measures a change in raw capacitance, instead of a change in resistance like prior works, to improve the quality of the output signal (Greenspan and Danielescu, 2020).

*Travel Distance:* The analog input is best suited for more precise movements, like the trigger on a video game controller. For this reason, we designed for a larger range of travel distances. By adjusting the height of the coil spring, we created analog keys with short (3.6 mm), medium (8.64 mm), and long (13.72 mm) travel distances. The activation force of the spring can also be modified by changing the coil thickness.

**DESIGN STEPS**

Leveraging the library of input keys and their customizable features, we created a QWERTY keyboard as an example.

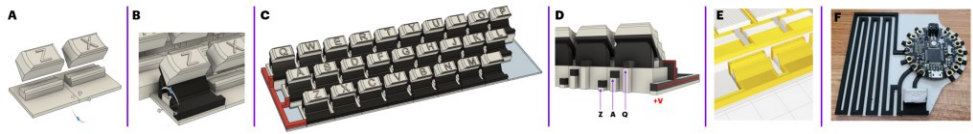

Figure 2. A) Constraining two non-conductive key parts next to one another B) Constraning the complementary condcutive part C) Full keybaord layout D) Return electrode routing to the rear of the input device E) Separate .STL files in slicing software F) 3D printed pull down resistor and socket for direct microcontroller inegration

We started by selecting the digital key with the short travel distance and high activation force from the library. An individual non-conductive part file was created for each letter of the alphabet plus a 'space' key, for a total of 27 inputs. Each key's complementary conductive and non-conductive components were saved, resulting in 27 non-conductive and 27 conductive part files, though all conductive part files were identical.

All non-conductive part files are assembled within Fusion 360, allowing the designer to place the keys in their desired location. The base of each key is constrained to its adjacent keys, resulting in the default key spacing mentioned earlier (Figure 2A). For this example, keys were placed in a QWERTY layout with the 'space' key located to the right of the lower 'M' key. The designer then inserts each conductive part into the assembly, and constrains them to their complementary non-conductive part (Figure 2B). The conductive traces connecting to the positive electrode extend to the left and right edge of each key, creating a continuous conductive row. An additional conductive trace is added that connects multiple conductive rows (Figure 2C, red trace), allowing for a single trace extending to the back of the device to supply the signal voltage. Each key includes a return electrode that travels to the back of the device. To accommodate multiple rows of keys, we added additional height to the non-conductive base so the signal from the return electrodes can travel beneath other rows without interfering (Figure 2D).

After all parts are imported into the assembly and constrained, the perimeter can be selected using the project geometry tool, and the base can be extended to create the desired shape. For the keyboard example, the left and right edges are extended, so the outer shell is a simple rectangle (Figure 2C, blue shaded region). The completed layout is exported as two separate stereolithography (.STL) files, one for the non-conductive and one for the conductive portions of the assembly. The files were imported into Cura where individual printing properties are assigned to each part (Figure 2E). For the keyboard shown, both parts were printed with a 0.2 mm layer height and a 90% infill density. We selected 90% to ensure the precise geometry of the conductive features wouldn't be affected by over extrusion. After the print finishes, the final step is connecting the electrodes on the back of the keyboard to a microcontroller. For the QWERTY keyboard we used an Arduino Mega 2560 Rev3 because of its large number of digital inputs, but for devices with fewer input keys a more common microcontroller like an Arduino Uno Rev3 can be used.

To further enhance the process, we designed and verified a 3D printed socket for direct integration with a microcontroller, the Adafruit FLORA (Figure 2F). This e-textile board has larger through hole connectors to accommodate a needle and conductive thread. The socket has conical electrodes that fit the FLORA using a small amount of Bare Conductive Electric Paint to secure the electrical connections. All electronic traces were printed with a cross section of 2.54 mm x 2.54 mm, with a measured resistance of ~200 ohm-cm. Resistors are created by adjusting trace length, such as the pull down resistor between the return electrode and ground in Figure 2F. Additionally, multiple digital keys can be read by a single analog input by varying the trace length from each key to the microcontroller.

## TECHNICAL EVALUATION

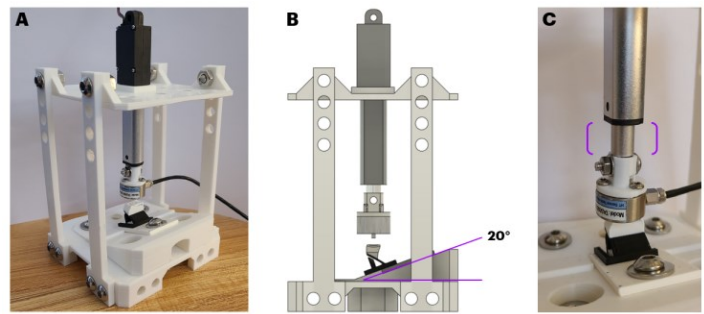

Figure 3. A) Custom 3D printed testing apparatus B) 20° adapter allowing normal keystroke activation direction C) End actuator position as keystroke has been made.

Physical input devices require a high level of durability and preliminary tests were completed to evaluate mechanical performance of printed keys. These tests also measured the activation force necessary to perform a keystroke for each of the travel distances and spring stiffness configurations which provides the physical parameters for the component library without needing to perform a finite element analysis (FEA). We tested and confirmed operation of the keys for all printed devices, including all 27 keys on the example QWERTY keyboard.

We created a custom testing apparatus to measure both the activation force and the durability of the keys (Figure 3A). The apparatus included a baseplate to mount keys and an elevated top surface where an ECO LLC 2" mini electric linear actuator was attached and positioned normal to the baseplate. At the distal end of the linear actuator, a 200 kg button style load cell, was affixed (HTC Model TAS606). The actuator was controlled by a microcontroller with a motor shield (Seeed 4A Motor Shield). Data from the load cell was fed to an amplifier and then read through a separate microcontroller to minimize noise (SparkFun Load Cell Amplifier - HX711). Keys were mounted to the testing apparatus at a 20° angle to match the natural angle of a user applying force (Figure 3B).

We performed a 1000 cycle deformation test on a digital key with a short travel distance and stiff spring. The linear actuator starts in the fully retracted position, then extends at a constant speed until it triggers the digital key (Figure 3C). The printed key functioned as an end stop switch, producing a signal that would retract the actuator and repeat the cycle. The digital switch showed no mechanical degradation during testing and completed 1000 successful consecutive keystrokes.

We also performed a 10 cycle test to measure the activation force of all digital key configurations. Due to limitations of the actuator, the linear actuator was manually extended to press the key to achieve a more precise reading from the load cell. The average activation force required for the short high stiffness spring was $2.51 \pm 0.15$ lbs, short low stiffness ($0.64 \pm 0.25$ lbs), medium high stiffness ($1.00 \pm 0.17$ lbs), medium low stiffness ($0.49 \pm 0.05$ lbs), long high stiffness ($1.46 \pm 0.13$ lbs) and long low stiffness ($0.97 \pm 0.08$ lbs). Measurements are used to characterize the forces in key presses as properties of 3D printed materials can vary drastically (Torres *et al.*, 2016; Cody *et al.*, 2019), making simulations inaccurate.

## APPLICATIONS

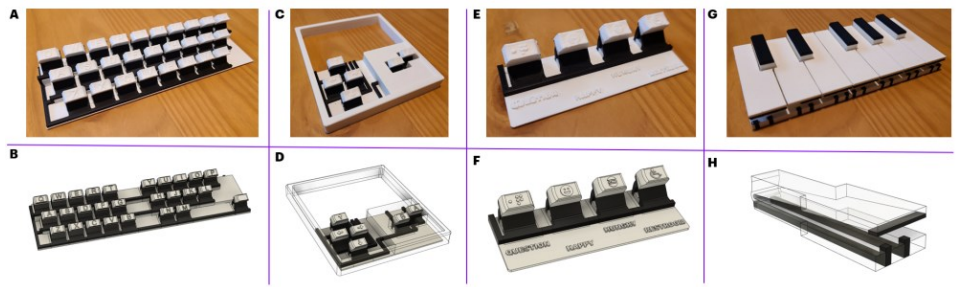

Figure 4. A) Standard QWERTY keyboard B) Custom spacing ergonomic keyboard C) Gamepad D) Wireframe view of gamepad conductive traces E) Custom AAC device F) CAD model of custom AAC device G) Piano keyboard H) Wireframe view of piano key

Beyond the keyboard example presented, this approach can be used to create a variety of input devices including an ergonomic keyboard, gamepad, augmentative and alternative communication (AAC) device and a piano keyboard. All devices were created using the same design approach and printed in a single print.

The printed QWERTY keyboard can be seen in (Figure 4A). This design was also modified for custom ergonomics with a new key spacing and layout, creating a gap between the left and right cluster of keys and a detached space key to demonstrate customization, (Figure 4B). This modified layout was created in the same amount of time as the standard keyboard, in contrast with the substantial design resources and manufacturing time and cost it would take to produce such a device utilizing traditional manufacturing methods. Beyond the keyboard form factor, we can arrange multiple keys in a custom layout to produce a handheld gamepad with a directional pad (D-Pad) and A & B buttons to mimic a standard controller (Figure 4C). In this application, we further demonstrate how the electronics can be printed into a final form factor by printing the full shell and leaving an opening in the top of the device where a display could be attached. We additionally show different ways to highlight or hide the conductive traces as seen in Figure 4D where the left-side is exposed and the right-side is embedded.

AAC devices are used by individuals with limited or no speech to communicate with others. There are multiple AAC devices available on the market, where each button triggers a different pre-recorded word or message (*Talkable IV*, 2017). The pre-recorded message can be changed, but the form factor of the device cannot. Using our approach, custom AAC devices can be printed to meet user's individual needs such as creating devices that can be integrated into the armrest of a wheelchair (Figure 4E). The process also enables custom labels that can be created based on user needs, including symbols or braille (Figure 4F).

Additional input key designs can also be added to the library. As an example, we created a piano keyboard, where the deformable components use a compliant mechanism design rather than a cantilever spring (Figure 4G). The thinness at the point of rotation allows the keys to deform, enabling three separate electrodes to connect when the key is depressed (Figure 4H). Just like our original digital key, the designer can adjust the piano key length, the required force and other interaction metrics. Once added to the part library, these keys can be combined with other keys and constrained to create a custom layout.

## CONCLUSIONS

The FlexKeys approach for creating tactile input devices is a significant step towards enabling a broader audience to design devices using 3D printing technologies. The input keys described provide an initial library, allowing users to choose their key types and customize multiple features such as travel distances, activation force, keycap textures and layout without the need for calculations, simulations, and verification. We demonstrated several complete input devices such as a keyboard with custom spacing, a gamepad, and AAC device as well as added an additional key design for a piano keyboard that can be created using this design approach. All designs have integrated electronic traces and can include a socket for direct microcontroller integration. Designs were put through preliminarily mechanical testing and evaluated for quality, connectivity, robustness, and activation force.

In this work we aimed to simplify the design and creation of tactile input devices and take the first step towards opening the design of physical input devices to a broader audience. Providing users with a process that uses customized 3D printed keys and produces final input devices that require no assemble or wiring can enable professionals like physical therapists, special education professionals and even avid gamers to create custom interface devices that meet the immediate needs of the end user. Additional steps are needed to realize this level of personal fabrication, such as creating custom scripts within Fusion 360 to facilitate design choices, a larger library of key designs and a method for automatic routing of electrical traces to a known microcontroller board. We present FlexKeys as an initial step to enable anyone to design and fabricate custom physical interface devices, including designs directly based on the assessment of individual user needs.